\documentstyle[11pt]{article}

\begin{document}

\thispagestyle{empty}

\begin{center}
{\Large {\bf Resistance to Collapse: The Tensor Interaction}}

\vspace{0.3in}

M. S. Fayache and L. Zamick

\begin{small}
{\it Department of Physics and Astronomy, Rutgers University,
	Piscataway, NJ 08855}
\end{small}

\end{center}

\begin{abstract}
We consider negative-parity excitations in $^4\mbox{He}$,
$J^{\pi}$=$0^-$ and $1^-$ with isospin $T$=$0$ and $T$=$1$. We show that,
in contrast to one-particle one-hole results, in a shell model
diagonalization none of the above collapses to zero energy as the
strength of the tensor interaction is increased.
\end{abstract}

\pagebreak

\section{Introduction}
Most of the studies of collapse of nucleon systems involve R.P.A.
calculations. One can trace the oncoming of the collapse by gradually
turning on an interaction to its full strength. Particle-hole
excitations with certain quantum numbers which originally have a high
excitation energy come down lower and lower in energy as the
interaction strength is increased, and eventually they come to zero
energy. At this point, it costs no energy to produce several of these
particle-hole states. The nature of the ground state changes
considerably.

Most of the interest in collapse has been for pion condensation. The
particle-hole states of interest have quantum numbers $S$=$1$ $T$=$1$ and
negative parity, e.g. $0^-$, $1^-$ spin dipole, $2^-$ etc. With the
interactions in current use, states with these quantum numbers are at
high energies. Also, consistently corresponding states with isospin
$T$=$0$ lie lower in energy than the $T$=$1$ states. But there is always
the underlying possibility, especially for the spin excitations, that
a stronger tensor interaction could help to bring down these modes in
energy.

In this work, we will not deal directly with the problem of pion
condensation but rather a related problem of increasing the overall
strength of the tensor interaction.
Rather than rely on R.P.A. calculations, we will perform the more
accurate shell model diagonalization. We intend to show that the
behavior of the energies of the 'particle-hole' states versus tensor
strength is quite different in the shell model diagonalization than it
is in the T.D.A. or R.P.A.

\section{Method:}

{\bf{ 2 (a) Lowest Order}}

In lowest order, the configuration for the ground state of
$^4\mbox{He}$ is $(0s)^4$, and the excited states mentioned above are
one-particle one-hole configurations. The $0^-$ state in lowest order
has the unique configuration $(0p\frac{1}{2}$ $0s\frac{1}{2}^{-1})$.
The energy of this state is given by
\begin{equation}
E(0^-,T)=\epsilon_ {0p_{\frac{1}{2}}} - \epsilon_{{0s_{\frac{1}{2}}}^{-1}} +
\langle (0p_{\frac{1}{2}}{0s_{\frac{1}{2}}}^{-1}) |V| (0p_{\frac{1}{2}}
{0s_{\frac{1}{2}}}^{-1}\rangle^{0^-,T}
\end{equation}
where $\epsilon_{i}$ are the single-particle energies and the last
term is the particle-hole interaction. It should be mentioned that we
calculate the single-particle energies with the same interaction that
is used to calculate the two-particle matrix elements. We used the
schematic (or democratic) interaction described in a work by Zheng and
Zamick \cite{ann}. It is of the form

\begin{equation}
V_{sche}=V_c+xV_{so}+yV_t
\end{equation}
where $c \equiv $ central, $s.o. \equiv $ spin-orbit, and $t \equiv$
tensor. For $x=1$, $y=1$ one gets a fairly good fit to the two-body
matrix elements of more realistic interactions like Bonn A. We focus
on the effects of the \underline{tensor} \underline{interaction} at
the energies of the states in $^4\mbox{He}$. We do this by varying
$y$, the strength of the tensor interaction. In the simple
one-particle one-hole picture, the single-particle energies do not
depend on $y$ (i.e. the first-order tensor contribution to these
energies is zero) and only the particle-hole matrix element is
affected. In the full shell model calculation the situation is more
complicated- there are many configurations.

In a $1p-1h$ calculation, the $J$=$0^-$ states have unique
configurations\\
$(0p_\frac{1}{2}{0s_\frac{1}{2}}^{-1})^{J=0,T}$
with $T$=$0$ or $1$.
Using equation (1), we note that as we increase $y$, the single-particle
energies $\epsilon$ do not change. Obviously, the particle-hole
interaction
$$V_{ph}=\langle (0p_{\frac{1}{2}}{0s_{\frac{1}{2}}}^{-1}) |V|
(0p_{\frac{1}{2}}{0s_{\frac{1}{2}}}^{-1}\rangle^{0^-,T}$$
will be linear in $y$. We find

$$V_{ph}(T=0)=2.575 - 3.820y\hspace{.2in} MeV$$

$$V_{ph}(T=1)=3.445 - 1.270y\hspace{.2in} MeV$$
Note that the coefficient of $y$ for $T$=$1$, i.e. the slope, is
$\frac{1}{3}$ that for $T$=$0$.
The excitation energy will decrease linearly in $y$ and we clearly can
get the $0^-$ states coming below the ground state by making $y$
sufficiently large. The $T$=$1$ state in this model is always higher
in energy than the $T$=$0$ $J$=$0^-$ states. We have not performed
R.P.A. calculations but we know from experience that when states come
down in energy in T.D.A. calculations they come down even faster in
R.P.A. calculations.

{\bf{ 2 (b) Matrix Diagonalization}}

We perform a shell model matrix diagonalization, using the OXBASH
code \cite{oxbash}, for the $J$=$0^+$ ground state and excited
$J$=$0^-$ and $1^-$ states with both isospin $T$=$0$ and $T$=$1$
in $^4\mbox{He}$. We allow the four nucleons to be anywhere in the
first three major shells $0s$, $0p$ and $1s$-$0d$.
We choose $^4\mbox{He}$ for practical reasons - it is easier to
perform a high-quality shell model calculation in such a light nucleus.

In Table I we give the shell model results as a function of the
strength parameter of the tensor interaction $y$ (for
$x$=$1$) for the excitation energies in $^4\mbox{He}$ of the lowest
$J$=$0^-$ and $1^-$ states with isospins $T$=$0$ and $T$=$1$. It
should be mentioned that we have removed the $1^-$  $T$=$0$ spurious
state- what is listed is the lowest non-spurious state. We also show
the behaviour of the excitation energies of the $J$=$0^-$ $T$=$0$ and
$T$=$1$ states as a function of $y$ in Fig 1.

For the $J$=$0^-$ states, we see that both the $T$=$0$ and $T$=$1$
states come down in energy as we increase $y$ up to a certain point
(i.e. $y$$\simeq$ $3$ for $T$=$0$) but then as $y$ is further increased
the excitation energy gets larger.

We note that the ground state binding energy changes as we increase $y$.
In lowest order, i.e $(0s)^4 $, there is no contribution
to the binding energy due to the tensor interaction (this holds for
any \underline{major} shell. However, the
nucleon-nucleon interaction induces configuration mixing into the
ground-state wave function. For this more complicated ground state, the
tensor interaction does contribute to the binding energy.

The change in binding energy in $MeV$ of the ground state relative to the case
$y$=$0$ is as follows:
\begin{tabbing}
\hspace{.5in}\= $y$=$0$\hspace{.3in} \=$0.00$ \hspace{1.in}
\= $y$=$1$\hspace{.3in} \=$1.83$   \\
\hspace{.5in}\> $y$=$2$\hspace{.3in} \> $7.01$\hspace{1.in}
\> $y$=$3$\hspace{.3in} \> $14.68$  \\
\hspace{.5in}\> $y$=$4$\hspace{.3in} \> $23.99$\hspace{1.in}
\> $y$=$5$\hspace{.3in} \> $34.42$  \\
\end{tabbing}
we see that this change of energy starts out quadratic in $y$ as we
would expect from a second-order tensor effect.

Although the full matrix diagonalization does not lead to the negative
parity excitations sinking below the ground state, the ground state
wave function does change as the tensor strength $y$ is increased. The
occupancy of shells higher than $0s$ increases with increasing $y$.
{}From $y=0$ to $y=5$, $0s$ occupancies are $3.53$, $3.46$, $3.28$,
$3.08$, $2.91$ and $2.76$. The corresponding $0p_\frac{1}{2}$
occupancies are $0.08$, $0.17$, $0.38$, $0.61$, $0.80$ and $0.95$.
Hence the nature of the ground state does change but it does so in a
continuous manner.

\section{Closing Remarks:}

Whereas simple $1p-1h$ calculations with very strong tensor
interactions can yield a collapse of negative parity excitations, we
find in superior full shell model calculations this is not the case.
We will not here go into a detailed analysis of why this is so.
Suffice to say that there are several second and higher order
corrections beyond those of the T.D.A. or R.P.A. First of all the
single particle energies get renormalized in second order \cite{x}
\cite{y}. Second the particle-hole interaction can get renormalized
e.g. by the exchange of a phonon between the particle and the hole
\cite{z}. The model we have presented here is not quite the same as
that of pion condensation. For example, we have multiplied the entire
tensor interaction by a constant $y$- this in effect contains effects
of both the $\rho$ and $\pi$ mesons. However, we feel that we have made an
important point- that many more effects than those contained in
$1p-1h$ T.D.A. or R.P.A. calculations must be taken into account.
\section*{Acknowledgment}
This work was supported by U.S. Department of Energy under Grant
DE-FG05-86ER-40299. We thank Dao-Chen Zheng for his help.

\pagebreak

\begin{small}

\end{small}

\pagebreak

\begin{small}

\noindent

{\bf Table I}.
The excitation energies of the lowest-lying negative parity states in
$^4\mbox{He}$ as a function of the strength of the tensor interaction
$y$.

\begin{center}
\begin{tabular}{|c|c|c|c|c|}\hline
$y$ & $J$=$0^-$ $T$=$0$ & $J$=$0^-$ $T$=$1$ & $J$=$1^-$ $T$=$0$ & $J$=$1^-$
$T$=$1$ \\ \hline
0. & 20.88&  22.80 & 19.98 & 21.35 \\

1. & 18.78&  22.06 & 21.75 & 22.16 \\

2. & 17.41&  22.40 & 24.79 & 23.70 \\

3. & 17.04&  24.09 & 28.98 & 25.95 \\

4. & 17.51&  26.77 & 33.88 & 28.79 \\

5. & 18.58&  30.13 & 39.20 & 32.07 \\

10. & 28.03&  51.97 & 67.27 & 51.59 \\ \hline
\end{tabular}
\end{center}
\end{small}


\begin{thebibliography}{99}
\bibitem{ann} D.C. Zheng and L. Zamick, Ann. Phys. (NY) {\bf 206}, 106(1991).
\bibitem{oxbash} A. Etchegoyen, W.D.M. Rae, N.S. Godwin, B.A. Brown,
	W.E. Ormand, and J.S. Winfield, the Oxford--Buenos Aires --- MSU
	Shell Model Code (OXBASH) (unpublished).
\bibitem{x} G.F. Bertsch and T.T.S. Kuo, Nucl. Phys. A112, 204(1968)
\bibitem{y} L. Zamick, D.C. Zheng, and H. M\"{u}ther,
	Phys. Rev. {\bf C 45}, 2763(1992).
\bibitem{z} D.C. Zheng, L. Zamick, M.S. Fayache and H. M\"{u}ther,
Annals of  Physics, in press.
\end{thebibliography}
\end{document}